\documentclass[10pt,pra,nofootinbib,notitlepage,twocolumn,superscriptaddress]{revtex4-1}

\usepackage[utf8]{inputenc}
\usepackage[T1]{fontenc}
\usepackage[english]{babel}  
\usepackage{float}
\usepackage{times}
\usepackage{dsfont}
\usepackage{hyperref,amsmath,amssymb,amscd}
\usepackage{amssymb,amsmath,amsfonts,amsthm}
\usepackage{mathtools}
\usepackage{verbatim}
\usepackage{enumerate}
\usepackage{graphicx}
\usepackage{pmat}
\graphicspath{{figs/}}
\usepackage{hyperref}
\usepackage{bbm}
\usepackage{booktabs}

\usepackage[caption=false]{subfig}

\usepackage{color}
\definecolor{dred}{rgb}{.8,0.2,.2}
\definecolor{ddred}{rgb}{.8,0.5,.5}
\definecolor{dblue}{rgb}{.2,0.2,.8}
\definecolor{dgreen}{rgb}{.2,0.5,.2}

\newcommand{\be}{\begin{equation}}
\newcommand{\ee}{\end{equation}}

\newcommand{\bea}{\begin{eqnarray}}
\newcommand{\eea}{\end{eqnarray}}

\begin{document}



\title{Experimental realization of noise-induced adiabaticity in nuclear magnetic resonance}
\author{Bi-Xue Wang}
\affiliation{State Key Laboratory of Low-Dimensional Quantum Physics and Department of Physics, Tsinghua University, Beijing 100084, China}
\affiliation{Tsinghua National Laboratory of Information Science and Technology, Beijing 100084, China}
\author{Tao Xin}
\affiliation{State Key Laboratory of Low-Dimensional Quantum Physics and Department of Physics, Tsinghua University, Beijing 100084, China}
\affiliation{Tsinghua National Laboratory of Information Science and Technology, Beijing 100084, China}
\author{Xiang-Yu Kong}
\affiliation{State Key Laboratory of Low-Dimensional Quantum Physics and Department of Physics, Tsinghua University, Beijing 100084, China}
\author{Shi-Jie Wei}
\affiliation{State Key Laboratory of Low-Dimensional Quantum Physics and Department of Physics, Tsinghua University, Beijing 100084, China}
\author{Dong Ruan}
\affiliation{State Key Laboratory of Low-Dimensional Quantum Physics and Department of Physics, Tsinghua University, Beijing 100084, China}
\author{Gui-Lu Long}
\thanks{gllong@mail.tsinghua.edu.cn}
\affiliation{State Key Laboratory of Low-Dimensional Quantum Physics and Department of Physics, Tsinghua University, Beijing 100084, China}
\affiliation{Tsinghua National Laboratory of Information Science and Technology, Beijing 100084, China}
\affiliation{The Innovative Center of Quantum Matter, Beijing 100084, China}
\date{\today}

\begin{abstract}
The adiabatic evolution is the dynamics of an instantaneous eigenstate of a slowly varing Hamiltonian. Recently, an interesting phenomenon shows up that white noises can enhance and even induce adiabaticity, which is in contrast to previous perception that environmental noises always modify and even ruin a designed adiabatic passage.  We experimentally realized a noise-induced adiabaticity in a nuclear magnetic resonance system. Adiabatic Hadamard gate and entangled state are demonstrated. The effect of noise on adiabaticity is experimentally exhibited and compared with the noise-free process. We utilized a noise-injected method, which can be applied to other quantum systems.
\end{abstract}

\maketitle
\section{Introduction}
\label{sec:Introduction}
The adiabatic principle is a fundamental concept in quantum mechanics. It states that a quantum system stays in its instantaneous eigenstate if the Hamiltonian is slowly varying~\cite{adiabatic1,adiabatic2}. Adiabatic process plays an important role in  quantum information processing and  quantum dynamics control, such as quantum adiabatic algorithm~\cite{QAA}, fault-tolerance against quantum errors~\cite{FT}, universal adiabatic and holonomic quantum computation~\cite{HQC,HQC2}, adiabatic passage~\cite{application1,application2,application3,application4}, adiabatic gate teleportation~\cite{application5}, and many other protocols~\cite{application6,application7,application8,application9,application10,application11,application12,application13}. In spite of extensive works in adiabaticity of closed systems~\cite{application1,application2,application3,application4,application10}, actual systems are open because of inevitable interactions between the systems and their surrounding environments~\cite{open1,open2}. Adiabaticity has theoretically been extended to  open quantum systems~\cite{application7,open3}. In particular, noise can be used to enhance the coherence and entanglement of quantum systems~\cite{noise1,noise2,noise3,noise4} and even induce adiabaticity~\cite{NIA1,NIA2}, rather than suppress the adiabatic passage. In Ref.~\cite{NIA1}, Jing \emph{et al.} derived a simple one-component dynamical equation governing the target instantaneous eigenstate and then they obtained an adiabatic condition when the integrand appearing in the integro-differential equation has a fast-varying factor. An external white dephasing noise can effectively induce the desired fast-varying factor to meet the adiabatic condition. It is called noise-induced adiabaticity.

In this work, we experimentally demonstrated adiabatic passage induced by white dephasing noise in nuclear magnetic resonance. Specifically, we first analyzed the adiabatic condition, and then experimentally implemented the adiabatic evolution of the time-dependent Hamiltonian with engineered white noise~\cite{ZHEN1,ZHEN2} to obtain the adiabatic Hadamard gate. In comparison, we also experimentally observe the noise-free evolution of the time-dependent Hamiltonian. Then we used the noise-induced adiabatic method to prepare the entangled state $\vert01\rangle+\vert10\rangle$ in an NMR system. Experimental results are consistent with numerical simulations.

This paper is organized as follows. In Sec.\ref{sec:condition}, we briefly review the condition and outcome of noise-induced adiabaticity. In Sec.\ref{sec:Hadamard gate}, we introduce our experiments of noise-induced adiabatic Hadamard gate, including details of the experimental process and results. In Sec.\ref{sec:entangled}, we report the preparation of  an entangled state using this noise-induced adiabatic method. Finally, a conclusion is given in Sec.\ref{sec:Conclusion}.

\section{the adiabatic condition}
\label{sec:condition}
Here we briefly give a summary of the adiabatic condition which was derived in Ref. \cite{NIA1} in full details. In general, using the Feshbach P-Q partitioning technique, the state and the effective Hamiltonian in the Schr$\ddot{\rm{o}}$dinger equation can be partitioned into
\begin{align}
\vert\psi(t)\rangle=
\begin{pmat}[{.}]
    P \cr\-
    Q \cr
  \end{pmat}
,   H=\begin{pmat}[{|}]
h & R \cr\-
W & D \cr
\end{pmat},
\end{align}
where $h$ and $D$ correspond to the self-Hamiltonians living in the subspace P and the subspace Q respectively, and R and W are
their mutual correlation terms. Consequently, we have
\begin{align}
\label{Eq.g}
\partial_tP(t)&=-ih(t)P(t)-\int_0^tds~g(t,s)P(s),\nonumber\\
        g(t,s)&=R(t)G(t,s)W(s).
\end{align}
We now rewrite the above Schr$\ddot{\rm{o}}$dinger equation into the adiabatic representation. The instantaneous eigenequation of $H(t)$ is
\begin{equation}
H(t)\vert E_n(t)\rangle=E_n(t)\vert E_n(t)\rangle,
\end{equation}
where $E_n(t)$'s and $\vert E_n(t)\rangle$'s are the instantaneous eigenvalues and nondegenerate eigenvectors, respectively. A state $\vert\psi(t)\rangle$ at time $t$ can then be expressed as
\begin{equation}
\vert\psi(t)\rangle=\sum_n\psi_n(t)e^{i\theta_n(t)}\vert E_n(t)\rangle,
\end{equation}
where $\theta_n(t)=-\int_0^tE_n(s)ds$ is the dynamical phase. Then we can obtain the following differential equation:
\begin{equation}
\label{5}
\partial_t\psi_m=-\langle E_m\vert\dot{E_m}\rangle\psi_m-\sum_{n\neq m}\langle E_m\vert\dot{E_n}\rangle e^{i(\theta_n-\theta_m)}\psi_n.
\end{equation}
Without loss of generality, applying Eq.(\ref{Eq.g}) and setting $P=\psi_0(t)$, we can get
\begin{equation}
\label{6}
\partial_t\psi_0=-\langle E_0\vert\dot{E_0}\rangle\psi_0-\int_0^tds~g(t,s)\psi_0(s).
\end{equation}
In this case, $R=[R_1,R_2,\dots]$ with $R_m=-i\langle E_0\vert\dot{E_m}\rangle e^{i(\theta_m-\theta_0)}$, and $W=R^{\dagger}$.  $G(t,s)=\mathcal{T}_{\leftarrow}\{\exp{ [-i\int_s^t D(s')ds'] }\}$ is a time-ordered evolution operator. The first term on the right-hand side of Eq.(\ref{6}) is similar to that in Eq.(\ref{5}), which corresponds to Berry's phase that can be switched off in a rotating frame~\cite{application4}.

With Eq.(\ref{6}), a general and crucial adiabatic condition can be derived into:
\begin{equation}
\label{7}
\int_0^t ds~g(t,s)\psi_0(s)=0.
\end{equation}
Obviously, the well-known adiabatic condition corresponds to the first-order approximation of the above result. The condition is satisfied when $g(t,s)=0$ or $g(t,s)$ is a rapid oscillating function~\cite{adiacondition}. Mathematically, it is easy to understand that the integral of the product of the fast-varying $g(t,s)$ and the slow-varying $\psi_0(s)$ leads to a vanishing result.

For a two-level system, when it is initially prepared at the eigenstate $\vert E_0\rangle$, the propagator $g(t,s)$~\cite{NIA1}is given by
\begin{equation}
\label{8}
g(t,s)=-\langle E_0(t)\vert\dot{E_1}(t)\rangle\langle E_1(s)\vert\dot{E_0}(s)\rangle e^{\int_s^t(iE-\langle E_1\vert\dot{E_1}\rangle)ds'},
\end{equation}
where $E=E_0-E_1$. If $E(t)$ can be manipulated by fast signal, the exponential term in $g(t,s)$ will play a crucial role to make the absolute value of the integral in Eq.(\ref{6}) or (\ref{7}) as small as possible. Later, Eq.(\ref{6}), (\ref{7}), and (\ref{8}) will be applied to analyze adiabatic condition.

\section{Noise-induced adiabatic Hadamard gate}
\label{sec:Hadamard gate}
We consider a time-dependent Hamiltonian describing adiabatic passage of single-qubit which can be written as
\begin{equation}
\label{9}
H(t)=J_0[a(t)\sigma_x+b(t)\sigma_z],
\end{equation}
where $a(t)=\frac{t}{T}$, $b(t)=1-\frac{t}{T}$ and $T$ is the whole adiabatic passage time. $H(0)= J_0\sigma_z$ , $H(T)= J_0\sigma_x$, the instantaneous eigenstate of $H(t)$ can be expressed as
\begin{equation}
\vert E_0(t)\rangle=\alpha\vert0\rangle+\beta\vert1\rangle,
\end{equation}
where $\alpha=\frac{k+b}{\sqrt{(k+b)^2+a^2}}$, $\beta=\frac{a}{\sqrt{(k+b)^2+a^2}}$, $k(t)=\sqrt{a(t)^2+b(t)^2}$. $\vert E_0(t)\rangle$ is the instantaneous eigenstate corresponding to eigenvalue $E_0(t)=J_0k(t)$.
The propagator (see appendix B for details) is
\begin{equation}
g(t,s)=\frac{e^{i\int_s^t E(s') ds'}}{4T^2k^2(t)k^2(s)}.
\end{equation}
When $T\to\infty$, the system could follow an adiabatic passage from an eigenstate $|0\rangle$ of $H(0)=J_0\sigma_z$ to $(|0\rangle+|1\rangle)/\sqrt{2}$ of $H(T)=J_0\sigma_x$. In addition, the adiabatic condition can be satisfied by injecting noise to the system. We replace the characteristic energy $J_0$ in the Eq.(\ref{9}) with $J_0+c(t)$. Here $c(t)$ is a white dephasing noise~\cite{ZHEN1,ZHEN2,ZHEN3}, written as
\begin{equation}
c(t)=\sum_{j=1}^N\alpha\sin(j\omega_0\ast t+\phi_j),
\end{equation}
where $\alpha$ is the noise amplitude and $\phi_j$ is the random phase. $N\omega_0$ determines the high-frequency cutoff $\omega_{\mathrm{cut}}$, with $\omega_0$ being the base frequency. We show that noise can render the general adiabatic condition valid since $g(t,s)$ becomes fast-varying function. Note that the noise term only rescales the eigenvalues $E_m$'s to $[1+c(t)/J_0]E_m$ but does not change their instantaneous eigenstates. Then we can apparently deduce that the imaginary part of $\alpha\beta^{\ast}$, denoted as Im($\alpha\beta^{\ast}$), is almost zero as a result of adiabaticity.

\begin{figure}[htbp]
\begin{center}
\includegraphics[width= 1\columnwidth]{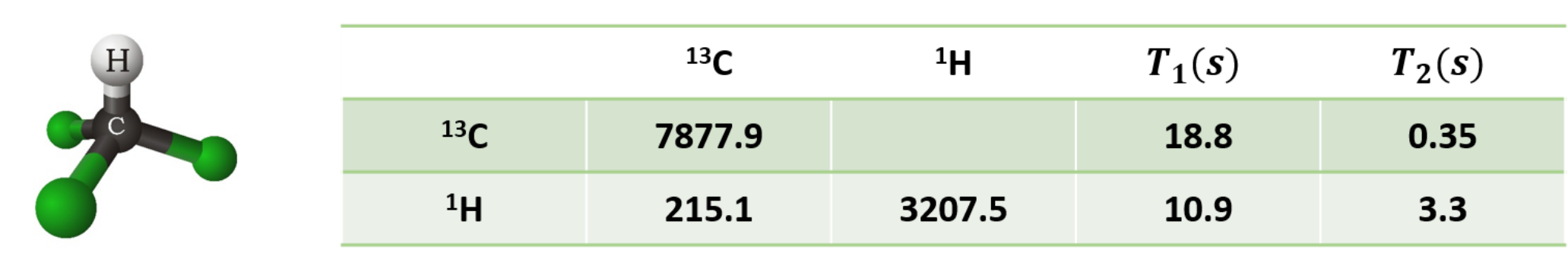}
\caption{\footnotesize{(coloronline) Molecular structure and relevant parameters of$~^{13}$C-labeled Chloroform. Diagonal elements in the table are the values of the chemical shifts (Hz) and offdiagonal element is the J-coupling constant (Hz) between$~^{13}$C and$~^{1}$H nuclei of the molecule. The longitudinal time T1 and transversal relaxation T2 are also provided in the right table.}}
\label{molecule}
\end{center}
\end{figure}

\begin{figure}[htbp]
\begin{center}
\includegraphics[width= 1\columnwidth]{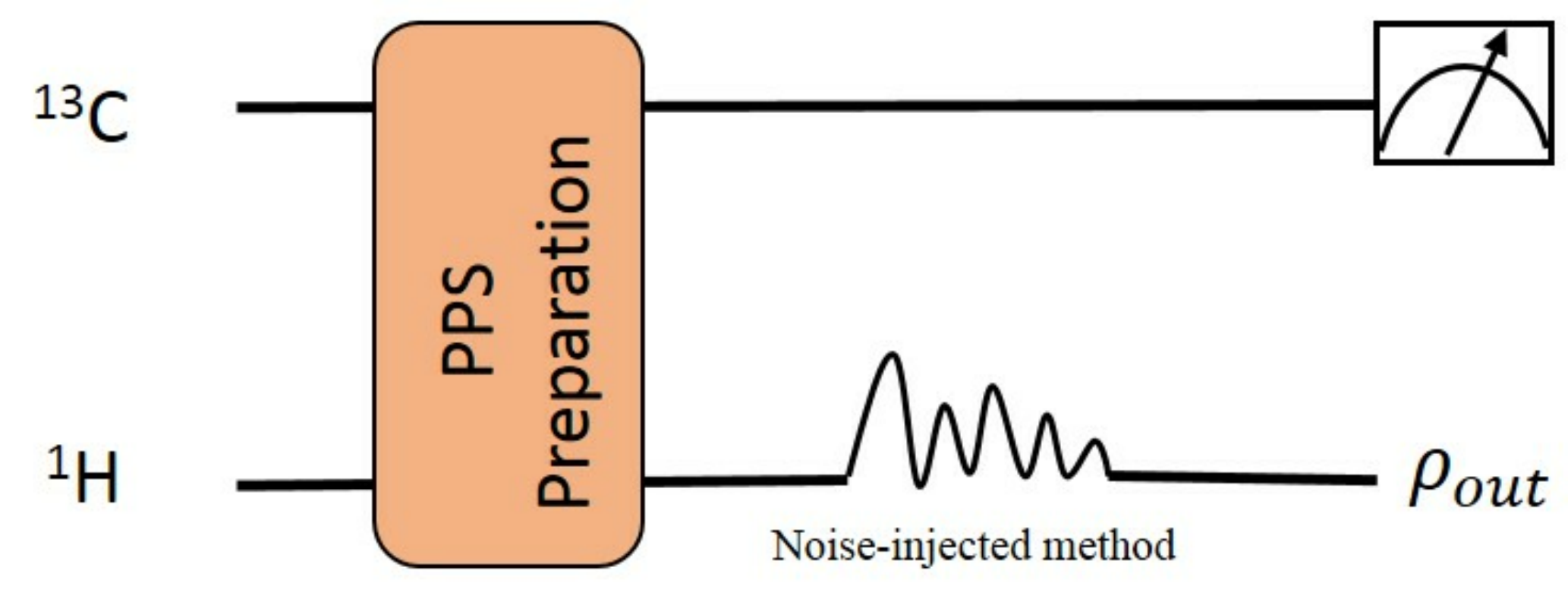}
\end{center}
\setlength{\abovecaptionskip}{-0.00cm}
\caption{\footnotesize{(coloronline) Implementation of the noise-induced adiabatic Hadamard gate in NMR system. The experiment consists of three steps: the initialization is to create a
two-qubit pseudopure state; Then evolution of time-dependent Hamiltonian is realized by noise-injected technique; Finally, observation of the C nucleus is to get the density matrix $\rho_{out}$ of H nucleus.}}
\label{circuit1}
\end{figure}

\begin{figure*}[htbp]
\begin{center}
\includegraphics[width= 2\columnwidth]{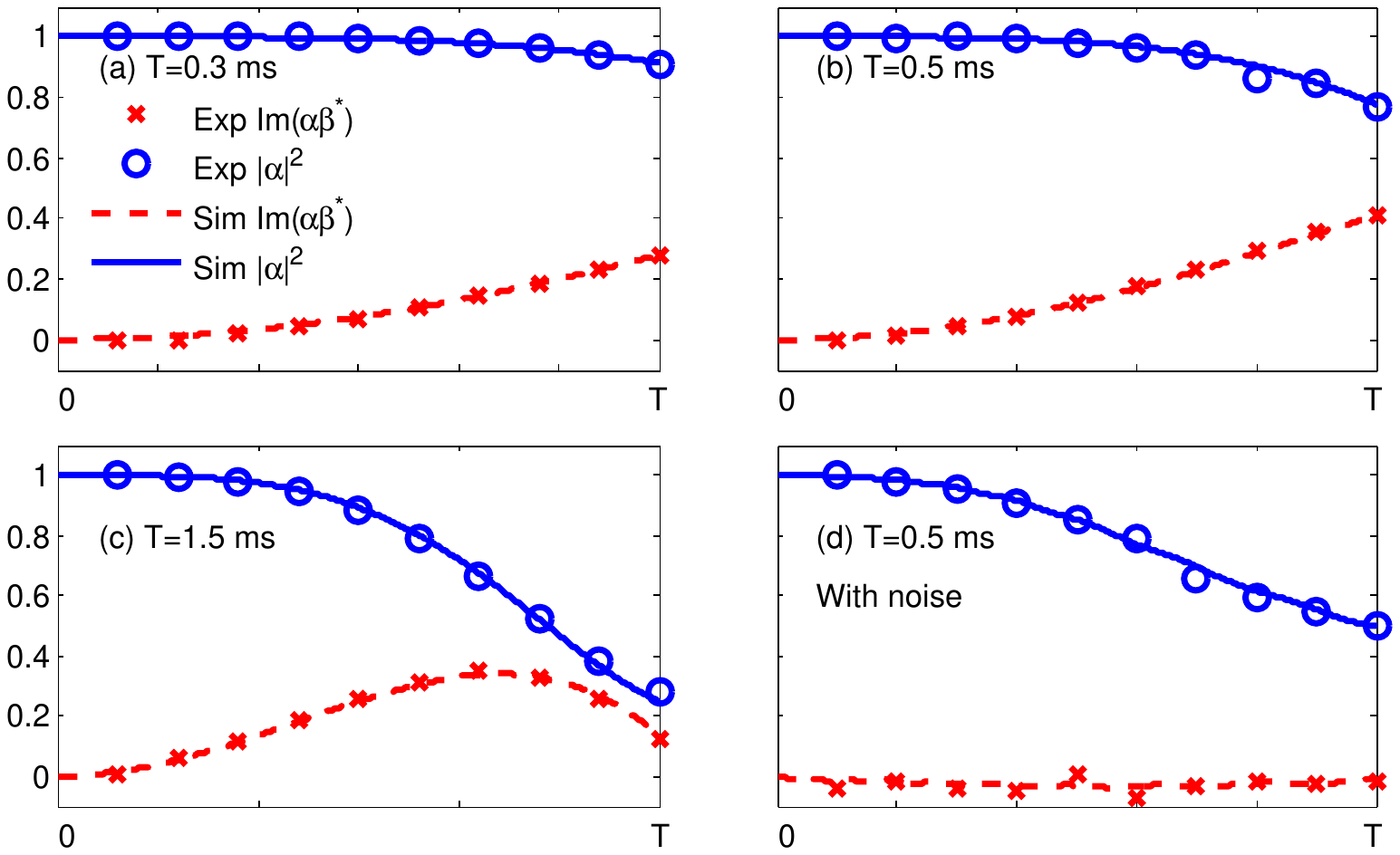}
\end{center}
\setlength{\abovecaptionskip}{-0.00cm}
\caption{\footnotesize{(coloronline) Experimental results for adiabatic Hadamard gate. The upper (blue) and lower(red) lines represent theoretical results of $\vert\alpha\vert^2$ and Im($\alpha\beta^{\ast}$), respectively. The (blue) circles and (red) crosses indicate experimental results of $\vert\alpha\vert^2$ and Im$(\alpha\beta^{\ast})$, respectively. (a) (b) (c) show results with noise-free evolution for the different total times $T=0.3$ms, $0.5$ms, $1.5$ms. And (d) shows noise-induced adiabaticity for $T=0.5$ms. Related noise parameters are $\alpha=4000$Hz, $\omega_{\mathrm{cut}}=5000$Hz, $\omega_0=1$Hz, $J_0=4000$Hz and the time step is $1\mu s$.}}
\label{onebit}
\end{figure*}

In order to demonstrate our scheme, all experiments were carried out on a Bruker 400MHz spectrometer at room temperature~\cite{NMR1,NMR2,NMR3,NMR4,NMR5,NMR6,NMR7}. We use the nuclear spins in a sample of 13C-labeled chloroform dissolved in deuterated acetone. Hence, the internal Hamiltonian of the system is
\begin{equation}
H_{\rm{int}}=\sum_{i=1}^2\omega_i\sigma_z^i+\frac{\pi J_{12}}{2}\sigma_z^1\sigma_z^2,
\end{equation}
where $\omega_i$ is the chemical shift of the $i$th nucleus and $J_{12}$ is the J-coupling constant between the nuclear spins. Fig.\ref{molecule} shows the molecular structure and properties of the sample. In Fig.\ref{circuit1}, we give the process of implementing the noise-induced adiabatic Hadamard gate where the H nucleus is the information carrier and the C nucleus is the observing qubit (see Appendix A).

Experiments are started from an initial thermal equilibrium state and we first generate a pseudopure state (PPS) using the spatial averaging technique ~\cite{PPS1,PPS2,PPS3} ,written as
\begin{equation}
\label{PPS}
\rho_0=\frac{1-\epsilon}{4}\mathcal{I}+\epsilon\vert00\rangle\langle00\vert,
\end{equation}
where $\epsilon\approx 10^{-5}$ and $\mathcal{I}$ is a $4\times4$ identity matrix. The first term of Eq.(\ref{PPS}) is neglected since the identity does not evolve under any unitary propagator and cannot be observed in NMR.

It is noticed that the Hamiltonian $H(t)$ in Eq.(\ref{9}) is similar to the Hamiltonian of hybrid noise, so it inspired us to make use of hybrid noise injecting technology to realize the evolution of time-dependent Hamiltonian~\cite{ZHEN1}. The simplest hybrid noise Hamiltonian for one qubit is $H(t)=\beta_z(t)\sigma_z+\beta_x(t)\sigma_x$. Specifically, the propagator is written as
\begin{equation}
U(t)=e^{-i\frac{\sigma_z}{2}\Delta\theta_t}\mathcal{T}e^{\{-i\int_{t_0}^t\beta_x(\tau)[\frac{\sigma_x}{2}\cos(\Delta\theta_{\tau})-\frac{\sigma_y}{2}\sin(\Delta\theta_{\tau})]d\tau\}},
\end{equation}
where $\Delta\theta_t=-i\int_{t_0}^t\beta_z(\tau)d\tau$.
In order to create a hybrid noisy environment, $\beta_x(t)$ and $\theta(t)$ are numerically generated with a desired noise power density spectrum and then used to modulate the corresponding continuous radio-frequency (RF) wave. It means that the continuous RF waves rotate the qubit around a changing axis in the equatorial plane and then a rotation of the $\Delta\theta_t$ angle around the z axis is applied at the end of the interval. We achieve the propagator $U(t)$ by modifying the amplitudes and phases of the RF wave. Let $\beta_x(t)=J_0a(t)$, $\beta_z(t)=J_0b(t)$, which are relevant in this case, $a(t)+b(t)=1$ and $\Delta\theta_t=\int_{t_0}^tJ_0(1-\frac{\tau}{T})d\tau$. We realize the time-dependent Hamiltonian by injecting hybird noise into the H qubit. Finally, We track the eigenstate $\vert E_0(t)\rangle=\alpha\vert0\rangle+\beta\vert1\rangle$, and then measure $\vert\alpha\vert^2$, which represents the probability of $\vert0\rangle$ state at $t$ time. Besides, we can check the value of the imaginary part of nondiagonal element Im($\alpha\beta^{\ast}$) to estimate whether it is adiabatic. In contrast to noise-free process, we solely substitute $J_0+c(t)$ for $J_0$ in the propagator $U(t)$ to accomplish evolution of time-dependent Hamiltonian with noise.

Here we consider noise-free process for the different total times $T=0.3$ms, $0.5$ms, $1.5$ms in Fig.\ref{onebit}(a)-(c). The final states at the end of the time $T$ are all not instantaneous eigenstate $(|0\rangle+|1\rangle)/\sqrt{2}$ and Im($\alpha\beta^{\ast}$) varies with time. Thus, adiabatic condition is not satisfied in the above scenario. Fig.\ref{onebit}(d) shows that $|\alpha|^2$ gradually decreases from 1 to 1/2 and meets with its instantaneous eigenstate at final time $T=0.5$ms. Im($\alpha\beta^{\ast}$) remains almost stable vanishing since the states always stay their instantaneous eigenstates in adiabatic passage. Experiments are almost consistent with theory. An important result is that white dephasing noises can even induce adiabaticity. In other words, we obtain the adiabatic Hadamard gate with the aid of noise.

\section{noise-induced adiabatic entangled state}
\label{sec:entangled}
Now we turn to two coupled two-level systems embedded in their individual baths, the time-dependent Hamiltonian is
\begin{equation}
\label{H2}
H(t)=J_0[a(\sigma_1^+\sigma_2^-+h.c.)+\omega(\sigma_1^z-\sigma_2^z)/4)],
\end{equation}
where $a=\frac{t}{T}$, $\omega=1-\frac{t}{T}$. If $\vert01\rangle$ and $\vert10\rangle$ are mapped into the two states for single qubit in Sec.\ref{sec:Hadamard gate}, namely $\vert01\rangle\rightarrow\vert0\rangle$,  $\vert10\rangle\rightarrow\vert1\rangle$. The propagator is
\begin{equation}
g(t,s)=\frac{e^{\int_s^t[J_0\mathcal{K}(s^{\ast})]ds^{\ast}}}{4T^2\mathcal{K}^2(t)\mathcal{K}^2(s)},
\end{equation}
where $\mathcal{K}(t)=\sqrt{T^2-2tT+2t^2}/T$. This model describes a finite time evolution defined by a period T. Similarly, When $T\to\infty$, the system could follow an adiabatic passage from an eigenstate $|01\rangle$ of $H(0)=J_0(\sigma_1^z-\sigma_2^z)/2$ to $(|01\rangle+|10\rangle)/\sqrt{2}$ of $H(T)=J_0[a(\sigma_1^+\sigma_2^-+h.c.)$. We also replace the characteristic energy $J_0$ in a Hamiltonian with $J_0 + c(t)$ to realize noise-induced adiabaticity.

\begin{figure}[htbp]
\setlength{\abovecaptionskip}{0.00cm}
\begin{center}
\includegraphics[width= 1.03\columnwidth]{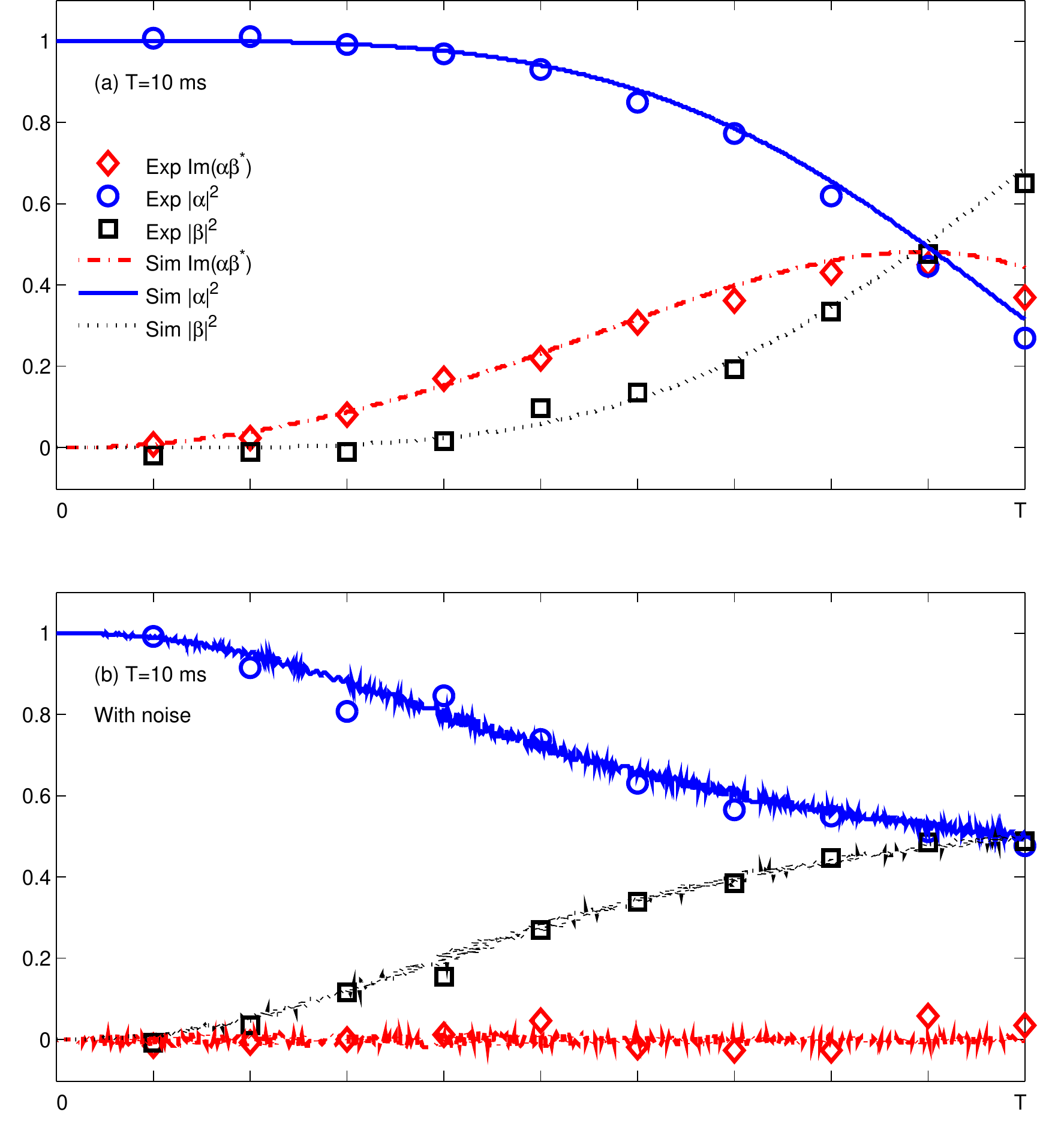}
\end{center}
\caption{\footnotesize{(coloronline) Experimental results for the preparation of adiabatic entangled state. The solid(blue) and dotted (black) lines represent theoretical results of $\vert\alpha\vert^2$ and $\vert\beta\vert^2$, respectively. The (blue) circles and (black) squares express experimental results of $\vert\alpha\vert^2$ and $\vert\beta\vert^2$, respectively. Theoretical and experimental results of Im($\alpha\beta^{\ast}$) are marked by dashed-dotted (red) lines and (red) diamond. (a) shows results of noise-free process for the time $T=10$ms. And (b) shows noise-induced adiabaticity for $T=10$ms. Related noise parameters are $\alpha=1000$Hz, $\omega_{\mathrm{cut}}=25000$Hz, $\omega_0=1$Hz, $J_0=100$Hz and the time step is $10\mu s$.}}
\label{twobit}
\end{figure}

We first prepare the initial state $|01\rangle$ by a $\pi$ rotation of H qubit along the x axis after PPS, then the evolution of Hamiltonian, shown in Eq.(\ref{H2}), is realized via the gradient ascent pulse engineering (GRAPE) technique~\cite{GRAPE1,GRAPE2}. The GRAPE approach provides over 99.5\% fidelity. The Fig.\ref{twobit} shows the instantaneous values of $|\alpha|^2$ and $|\beta|^2$ representing the populations of $|01\rangle$ and $|10\rangle$ respectively. For noise-free process, we observe that $|\alpha|^2=0.32,~|\beta|^2=0.68$ from Fig.\ref{twobit}(a) and the instantaneous state is not its eigenstate at the instantaneous time $T=10$ms. In Fig.\ref{twobit}(b), when the noise is added into the system, $|\alpha|^2=|\beta|^2\approx0.5$ and the instantaneous state is its eigenstate $(|01\rangle+|10\rangle)/\sqrt{2}$ at final time $T$. Im$(\alpha\beta^{\ast})$ almost keeps zero fixed with noise. It can be said that noise can induce adiabatic entangled state.

\section{Conclusion}
\label{sec:Conclusion}
We focus on an interesting phenomenon, that noise will not destroy the adiabatic process, but induce the adiabatic process contrary to our common sense. We experimentally demonstrate that the injection of addition white noise will accelerate the adiabatic process. Our experimental results are consistent with the theoretical simulations in the single-qubit and two-qubit NMR systems respectively, which supports our statement that the noise-induced adiabaticity can be realized experimentally. Nevertheless, adiabaticity can not be realized in unabiding finite time without noise. It is significant to be applied to many physical implementations of quantum information and quantum computing protocols such as holonomic and adiabatic quantum computing and the fast energy transfer. Furthermore, noise injected technique, which is applied experimentally to achieve evolution of time-dependent Hamiltonian in NMR system, can also be used for other quantum systems.

\begin{acknowledgments}
We thank X. L. Zhen for useful discussions. We are grateful to the following funding sources: National Natural Science Foundation of China under Grants No. 11774197 and No.61727801; National Basic Research Program of China under Grant No. 2015CB921002.
\end{acknowledgments}

\section*{Appendix A: Error Analysis}
\label{Error Analysis}
\begin{figure}[htbp]
\begin{center}
\includegraphics[width= 1\columnwidth]{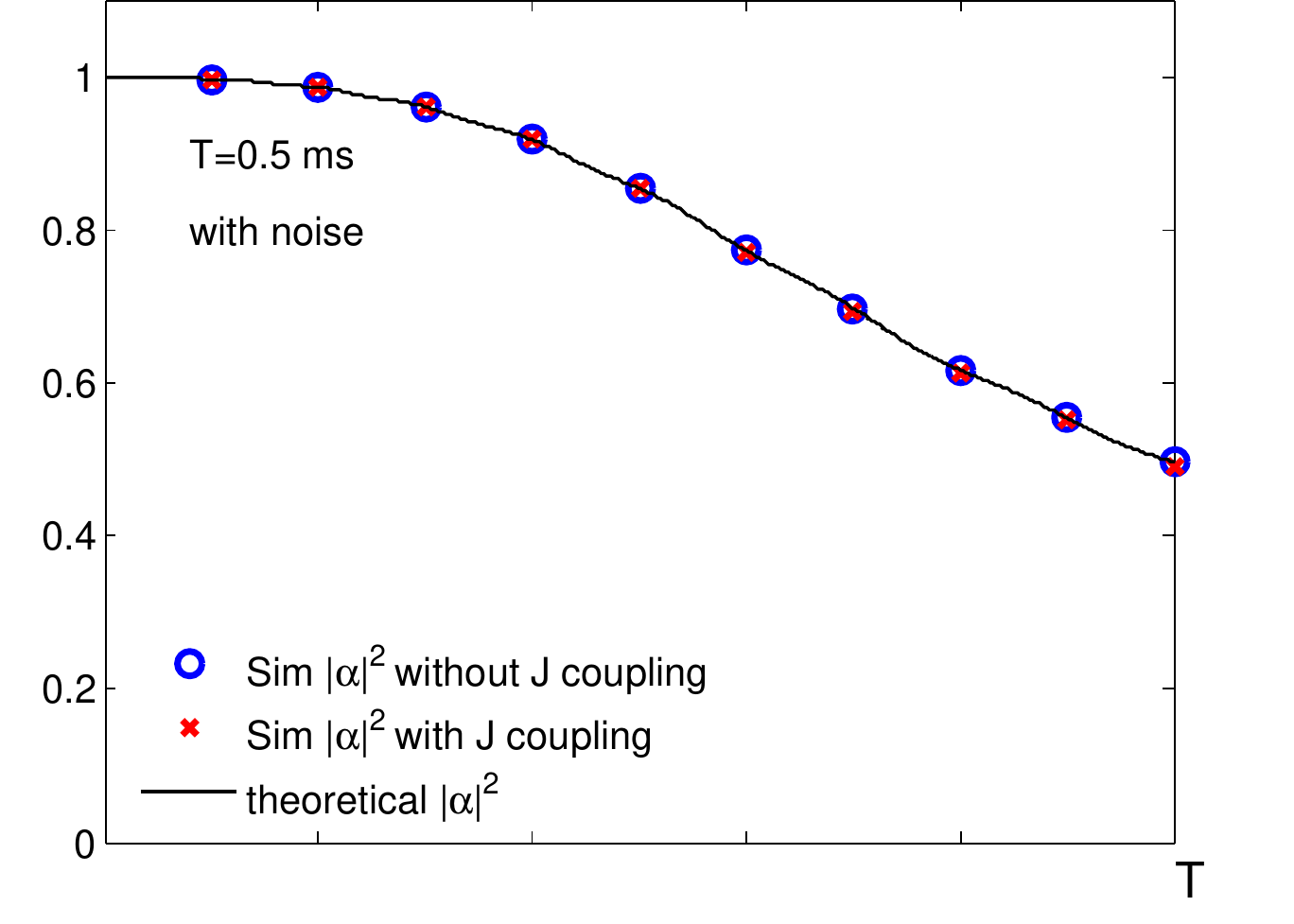}
\end{center}
\setlength{\abovecaptionskip}{-0.00cm}
\caption{\footnotesize{(coloronline) A comparison of the two simulation results about $J$ coupling. The (blue) circles or (red) crosses represent simulation results of $\vert\alpha\vert^2$ without or with $J$ coupling. The solid (black) line is theoretical results. The relevant experimental parameters are from Fig.\ref{onebit}(d).}}
\label{Jcoupling}
\end{figure}
$J$ coupling interaction exists between hydrogen nuclei and carbon nuclei. To show that the coupling barely has any effect on our experiments, we have done two numerical simulation with different Hamiltonians. One of them contains the $J$ coupling interaction term in the two-qubit system and the other one is for the single-qubit system without $J$ coupling. A comparison of the simulation results is shown in Fig.\ref{Jcoupling}, which nearly consists with theory. The relative errors between the two simulation results are less than 1\%. Hence it indicates that the existence of carbon nuclei makes no difference in our experiments.

\section*{Appendix B: $~\textbf{g~(t, s)}$ in details}
In sec.\ref{sec:Hadamard gate}, $H(t)=J_0[a(t)\sigma_x+b(t)\sigma_z]$, then we get obtain the instantaneous eigenstates
\begin{align}
\vert E_0(t)\rangle&=\frac{b+k}{\sqrt{(b+k)^2+a^2}}\vert0\rangle+\frac{a}{\sqrt{(b+k)^2+a^2}}\vert1\rangle, \label{18}\\
\vert E_1(t)\rangle&=\frac{b-k}{\sqrt{(b-k)^2+a^2}}\vert0\rangle+\frac{a}{\sqrt{(b-k)^2+a^2}}\vert1\rangle, \label{19}
\end{align}
where $a(t)=\frac{t}{T}$,  $b(t)=1-\frac{t}{T}$, $k(t)=\sqrt{a(t)^2+b(t)^2}$. Corresponding eigenvalues are $E_0(t)=J_0k(t)$ and $E_1(t)=-J_0k(t)$. According to Eq.(\ref{18}) (\ref{19}),
\begin{align}
\langle E_0(t)\vert\dot{E_1}(t)\rangle&=\frac{\langle E_0\vert\dot{H}(t)\vert E_1(t)}{E_1-E_0}=\frac{J_0}{Tk(t)E(t)}, \\
\langle E_1(t)\vert\dot{E_0}(t)\rangle&=\frac{\langle E_1\vert\dot{H}(t)\vert E_0(t)}{E_0-E_1}=-\frac{J_0}{Tk(t)E(t)}, \\
\langle E_1(t)\vert\dot{E_1}(t)\rangle&=0.
\end{align}
where $E(t)=E_1(t)-E_0(t)=-2J_0k(t)$. Substituting above equations into Eq.(\ref{8}), we can calculate
\begin{align}
g(t,s)&=-\langle E_0(t)\vert\dot{E_1}(t)\rangle\langle E_1(s)\vert\dot{E_0}(s)\rangle e^{\int_s^t(iE-\langle E_1\vert\dot{E_1}\rangle)ds'}  \nonumber \\
        &=-[-\frac{1}{2Tk^2(t)}][\frac{1}{2Tk^2(s)}] e^{\int_s^t iE(s') ds' }\nonumber\\
        &=\frac{\exp{[i\int_s^t E(s') ds']}}{4T^2k^2(t)k^2(s)}.
\end{align}

\end{document}